\documentclass[%
 reprint,
superscriptaddress,
 amsmath,amssymb,
 aps,
pra,
]{revtex4-2}
\usepackage{graphicx}
\usepackage{dcolumn}
\usepackage{bm}
\usepackage{physics}
\usepackage{hyperref}
\usepackage[caption=false]{subfig}
\usepackage{booktabs}
\usepackage{moreverb}
\usepackage{color}

\newcommand{\mE}{\mathcal{E}}
\DeclareMathOperator*{\argmin}{arg\,min} 

\immediate\write18{texcount -inc -incbib 
-sum borra.tex > /tmp/wordcount.tex}

\immediate\write18{texcount -char -freq
 borra.tex > /tmp/charcount.tex}

\usepackage{amsthm}

\theoremstyle{plain}

\newtheorem*{theorem*}{Theorem}
\newtheorem{theorem}{Theorem}

\usepackage{float}
\usepackage[ruled,linesnumbered]{algorithm2e}

\begin{document}

\preprint{arXiv}

\title{Informationally complete POVM-based shadow tomography}


\author{Atithi Acharya}
\affiliation{\footnotesize Department of Physics and Astronomy, Rutgers University, Piscataway, NJ 08854, USA}
\author{Siddhartha Saha}
\affiliation{\footnotesize Department of Physics and Astronomy, Rutgers University, Piscataway, NJ 08854, USA}
\author{Anirvan M. Sengupta}
\affiliation{\footnotesize Department of Physics and Astronomy, Rutgers University, Piscataway, NJ 08854, USA}
\affiliation{Center for Computational Mathematics and Center for Computational Quantum Physics, Flatiron Institute, New York, NY 10010 USA}


\begin{abstract}
Abstract: Recently introduced shadow tomography protocols use `classical shadows’ of quantum states to predict many target functions of an unknown quantum state. Unlike full quantum state tomography, shadow tomography does not insist on accurate recovery of the density matrix for high rank mixed states. Yet, such a protocol makes multiple accurate predictions with high confidence, based on a moderate number of quantum measurements. One particular influential algorithm, proposed by Huang, Kueng, and Preskill \cite{Huang_2020}, requires additional circuits for performing certain random unitary transformations. In this paper, we avoid these transformations but employ an arbitrary informationally complete POVM and show that such a procedure can compute $k$-bit correlation functions for quantum states reliably. We also show that, for this application, we do not need the median of means procedure of Huang {\it et~al}. Finally, we discuss the contrast between the  computation of correlation functions and fidelity of reconstruction of low rank density matrices. 
\end{abstract}



\maketitle


\section{Introduction}
Recent advances in quantum information processing often require characterizing quantum states prepared during various stages of a procedure. As a result, the problem of characterising a quantum state, more specifically, a density matrix, from measurements on an ensemble of identical states, known as quantum state tomography (QST), has seen a surge of interest \cite{Torlai_2018,Carrasquilla_2019, Huang_2020}. One of the key challenges is that, for $n$-qubit quantum systems, the density matrix is of size $2^n\times 2^n$. As the number of qubits become large,  inferring the density matrix from a  limited number of measurements becomes difficult.


Can we get away without fully characterizing the quantum state, but by constructing an approximate classical description that predicts many different functions of the state accurately?  Shadow Tomography \cite{aaronson2018shadow} precisely aims to do this, namely, predict a power law number of observations in number of qubits, $n$, from $O(n)$ copies of the density matrix $\rho$. This idea was taken further by Huang {\it et~al.\@}~\cite{Huang_2020} who have constructed such a description of low sample complexity via classical shadows ($\hat{\rho}$), related to states without any entanglement in the appropriate basis, corresponding to each copy of $\rho$ .

Quantum measurement requires specifying a set of Positive Operator Valued Measures (POVMs) \cite{Nielsen} which is a generalization of a complete set of projection operators. The work by Huang {\it et~al.\@}~\cite{Huang_2020} involves measurements via projection operators. Since projection operators are not informationally complete (see Sec.~\ref{sec:IC}),  Huang {\it et~al.\@}~employ a set of random unitary transformations before taking measurements. In this work, we directly employ a complete or overcomplete POVM system and perform shadow tomography. We also gain some insight into how the choice of POVM affects the efficiency of the method.

\section{Generalized Measurements}
\label{sec:POVM}
A projective measurement is described by an observable, $A$, a Hermitian operator on the state space of the system being observed. The observable has a spectral decomposition, $ A=\sum_{a}aP_{a}$
where $P_{a}$ is the projector onto the eigenspace of $A$ with eigenvalue $a$. The possible outcomes of the measurement corresponding to the eigenvalues, $a$, of the observable and the outcome probability is $p(a)=\bra{\psi}P_{a}\ket{\psi}$.
Projection Valued Measures (PVMs) are a special case of general measurements, where the measurement operators are Hermitian and orthogonal projectors. A set of Positive Operator Valued
Measures (POVMs) \cite{Nielsen} forms a generalization of PVMs. The index $a$ in the POVM element $M_{a}$ refers to the measurement outcomes that may occur in the experiment. The probability of the measurement outcome is given by $ \label{eq:Born}
p(a)= tr(\rho M_{a})$ and the post measurement density matrix can be written as $\frac{K_{a}\rho K^{\dagger}_{a}}{tr(K_{a}\rho K^{\dagger}_{a})}$, where $\{K_{a}\}$ are the Kraus operators \cite{Nielsen} corresponding to the POVM, with $K^{\dagger}_{a}K_{a}=M_{a}$.
The operators $\{ M_{a}\}$ form a complete set of Hermitian non-negative operators. Namely, they satisfy $Hermiticity: M_{a}=M^{\dagger}_{a}$, $Positivity: \bra{\psi}M_{a}\ket{\psi} \geq 0$ for any vector $\ket{\psi}$ and $Completeness: \sum_{a}M_{a}=\mathbb{I}$. Such a POVM could be thought of as a partition of unity by non-negative operators.


\subsection{Informational completeness}
\label{sec:IC}

The density matrix, $\rho$, is a Hermitian and unit trace operator. If we have a $d$-dimensional system, $\rho$ will be a complex square matrix represented by $d^2-1$ real parameters. The operator space for this $d$-dimensional operator will be however spanned by $d^2$ linearly independent basis operators. Note that PVMs only have $d$ projection operators. They are capable of providing only the diagonal elements of $\rho$ in a particular orthonormal basis, leaving out potential entanglement-related information from the off-diagonal elements. Thus, PVMs are examples of POVMs that are informationally undercomplete.

If the number of outcomes $k$ satisfies $k\geq d^2$, and we can form exactly $d^2$ linearly independent operators by linearly combining the set of POVMs, such POVMs will be called informationally complete (IC). However, in the most common terminology, informationally complete actually refers to the minimally complete POVM ($k=d^2$). If we proceed to reconstruct the density matrix for an informationally complete POVM, we can expand $\rho$ as 
\begin{equation}
\rho = \sum^{d^2-1}_{a=0}\xi_{a}M_{a}. 
\label{eq:expand}
\end{equation} 

If $k=d^{2}$ we have a informationally complete (\textbf{minimally complete}) basis set. However, if we have $k>d^2$ $\implies$ it forms an informationally \textbf{overcomplete} set \cite{Renes_2004}. IC POVM has been used for entanglement detection \cite{dimic2018single} and for individual elements of the density matrix \cite{morris2019selective}.

We start out by giving the example of a rather simple overcomplete set in the single-qubit Hilbert space, $\boldsymbol{M}_{\text{Pauli}-6}$. Pauli-6 POVM has 6 outcomes  $\boldsymbol{M}_{\text{Pauli}-6}=\big\{M_{0} = \frac{1}{3}\times\ketbra{0}{0},
M_{1}= \frac{1}{3}\times\ketbra{1}{1}, M_{2}= \frac{1}{3}\times\ketbra{+}{+}, 
M_{3}= \frac{1}{3}\times\ketbra{-}{-}, 
M_{4}= \frac{1}{3}\times\ketbra{l}{l}, 
M_{5}= \frac{1}{3}\times\ketbra{r}{r}  $ where $\{\ket{0}, \ket{1}\}$, $\{\ket{+}, \ket{-}\}$, and $\{\ket{r},\ket{l}\}$  
stand for the eigenbases of the Pauli operators $\sigma^{z}$, $\sigma^{x}$, and $\sigma^y$, respectively. Experimentally, it can be implemented directly by first randomly choosing $x$, $y$, or $z$, and
then measuring the respective Pauli operator, which justifies the $1/3$ factor. However, other probabilities will also be valid for this example of an overcomplete POVM.

Now, let us give an example of a minimally complete POVM, the Pauli-4 POVM:$\,\,\boldsymbol{M}_{\text{Pauli}-4}=\big\{M_{0} = \frac{1}{3}\times\ketbra{0}{0},
M_{1}= \frac{1}{3}\times\ketbra{+}{+}, M_{2}= \frac{1}{3}\times\ketbra{l}{l}, 
M_{3}= \frac{1}{3}\times\big(\ketbra{1}{1}+\ketbra{-}{-}+\ketbra{r}{r}\big)\big\}$. As a sanity check for the completeness relation, one can see 
$\sum_{a}M_{a}=1/3(\ketbra{0}{0}+\ketbra{1}{1}+ \ketbra{+}{+} + \ketbra{-}{-}+ \ketbra{l}{l}+ \ketbra{r}{r})=\mathbb{I}$. The experimental procedure will be similar to that of Pauli-6 POVM, with an additional step where three different outcomes of Pauli-6 are identified as the single element of Pauli-4, $M_3$. Thus, this set contains an element which is not a rank-1 projector.\\
The third one is the tetrahedral POVM $\boldsymbol{M}_{\text{tetra}}=\big\{M_a=\frac{1}{4}(\mathbb{I}+\boldsymbol{s}_a\cdot\boldsymbol{\sigma})\big\}_{a\in\{0,1,2,3\}}$, whose outcomes correspond to sub-normalized rank-1 projectors along the directions 
$\boldsymbol{s}_0=(0,0,1)$, $\boldsymbol{s}_1=(\frac{2\sqrt{2}}{3},0,-\frac{1}{3})$,  
$\boldsymbol{s}_2=(-\frac{\sqrt{2}}{3},\sqrt{\frac{2}{3}},-\frac{1}{3})$, 
and $\boldsymbol{s}_3=(-\frac{\sqrt{2}}{3},-\sqrt{\frac{2}{3}},-\frac{1}{3})$ in the Bloch sphere. Since the tetrahedron formed is regular, it forms an example of a symmetric informationally complete (SIC) POVM. The experimental implementation 
of $\boldsymbol{M}_{\text{tetra}}$ relies on Neumark's dilation theorem. The theorem implies that $\boldsymbol{M}_{\text{tetra}}$ 
can be physically realized by coupling the system qubit to an ancillary qubit and performing 
a von Neumann measurement on the two qubits (see Ref. \cite{Carrasquilla_2019, PhysRevA.86.062107} for explicit constructions).

\section{Classical Shadows with POVM\lowercase{s}}
\label{sec:Classical_Shadows}
Aaronson introduced the idea  of ``pretty good tomography"\cite{Aaronson_2007}, with the focus on predicting many observations accurately, based on $N$ copies of the density matrix. This idea parallels the ``learnability" of quantum states in a Probably Approximately Correct (PAC) sense \cite{PAC}. Proceeding along this line, he later introduced the concept of Shadow Tomography  \cite{aaronson2018shadow}, where from $N$ copies of the density matrix $\rho$, we want to predict   $L$ different linear target functions $tr(O_{1}\rho), tr(O_{2}\rho).\ldots tr(O_{L}\rho)$ up to an additive error less than $\epsilon$.

Huang {\it et~al.\@} \cite{Huang_2020} build their methods on the idea of Shadow Tomography \cite{aaronson2018shadow}. They repeatedly perform a measurement procedure, i.e.~apply a random unitary to rotate the state ($\rho \mapsto U \rho U^\dagger$) and perform a computational-basis measurement. 
Then, after the measurement, they apply the inverse of $U$ to the resulting computational basis state. This procedure collapses $\rho$ to a snapshot $U^\dagger |\hat{b} \rangle \! \langle \hat{b}|U$,
producing a quantum channel $\mathcal{M}$, which depends on the ensemble of (random) unitary transformations.

If the collection of unitaries is defined to be tomographically complete, namely, if the condition i.e.\ for each $\sigma \neq \rho$, there exist $U \in \mathcal{U}$ and $b$ such that $\langle b| U \sigma U^\dagger |b \rangle \neq \langle b| U \rho U^\dagger |b \rangle \label{condition_tc}$ is met, then
 $\mathcal{M}$ --- viewed as a linear map --- has a unique inverse $\mathcal{M}^{-1}$.  Huang {\it et~al.\@}~\cite{Huang_2020} set
\begin{align}
\hat{\rho} = \mathcal{M}^{-1} \left( U^\dagger | \hat{b} \rangle \! \langle \hat{b}|U \right) && \text{(classical shadow)}.
\label{eq:classical-shadow-appendix}
\end{align}

Although the inverted channel $\mathcal{M}^{-1}$ is not physical (it is not completely positive), one can still apply $\mathcal{M}^{-1}$ to the (classically stored) measurement outcome $U^\dagger |\hat{b}\rangle\!\langle\hat{b}| U$
 in a completely classical post-processing step.  Even if an individual sample of $\hat{\rho}$ is not a density matrix, the expectation of $\hat{\rho}$'s is the original density matrix $\rho$. One can use this property to get a good prediction of measurements performed on $\rho$.


If, instead of working with the computational basis measurements, we decide to use an IC POVM (Sec. \ref{sec:IC}), we can avoid dealing with particular random unitary ensembles. The only thing we need to make sure is that the resulting channel $\mathcal{M}$ is invertible.


\subsection{POVMs for the the $n$-qubit system}
From single qubit POVMs $\{M_{a}\}$, we introduce $k^n$ operators by taking tensor products and form POVMs for the $n$-qubit system:
 $\boldsymbol{M}=\big\{M_{a_1}\otimes M_{a_2}\otimes..M_{a_n}\big\}_{a_1, \hdots a_n}$.
The outcomes of this measurements in this system are of the form 
$\vec{\mathbf{a}}=( a_{1}, a_{2},....a_{n} )$. Now, we discuss how to form shadows from such an observation.
 
\subsection{ A synthetic measurement channel}
\label{sec:Measurement_channel} Let the POVM elements be diagonalised as follows:
$M_{a}=\sum_{i}\lambda^{a}_{i}\ketbra{i,a}{i,a}, \quad  \forall i,a, \lambda^{a}_{i} \geq 0 $, since $M_{a} \succeq 0$. Let $f: \mathcal{R}^{+}_0 \to \mathcal{R}^{+}_0 $ be a strictly monotonic function which will be applied to the eigenvalues of the POVM elements. The function $f$ is defined on $\mathcal{R}^{+}_0$ since the eigenvalues are non-negative.
 The probability outcome `$a$' is given as
\begin{equation}\label{eq:Born_povm}
p(a)= tr(\rho M_{a}).\end{equation}
Each time we perform a measurement and get an outcome `$a$', we construct a pure output state  $\ketbra{i,a}{i,a}$ with probability $p(i|a)=\frac{f(\lambda^{a}_{i})}{\sum_{j}f(\lambda^{a}_{j})}$. We assume each $M_{a}$ to be non-zero, guaranteeing that the denominator $\sum_{j}f(\lambda^{a}_{j})\neq 0$. Although this is a synthetic channel, we will refer to it as the measurement channel, in analogy with the case where $\{M_a\}$ are projections. 

The measurement channel, for a single qubit, can be defined as \begin{equation}
    \label{eq:meas_channel_general}
    \Tilde{\rho}=\mathcal{M}(\rho)=\sum_{a}p(a)\sum_{i}p(i|a)\ketbra{i,a}{i,a}.
\end{equation}

For simplicity, in the following discussion, we consider the case where the highest eigenvalue of each $M_a$ is non-degenerate. The modifications needed for the general case are obvious.
If a particular POVM element is not a rank one projector and the function $f$ is very steeply increasing, then the overwhelmingly likely output is  $\ket{\psi_{a}}\bra{\psi_{a}}$,  where $\ket{\psi_{a}}$ is the eigenvector corresponding to the highest eigenvalue of $M_{a}$. An example of such a function is $f(\lambda)=\lambda^{m}$ in the large $m$ limit. 
In the large $m$ limit, as we perform a measurement, the output is $\ketbra{\psi_{a}}{\psi_{a}}$ (snapshots) with probability $tr(\rho M_a)$. The measurement channel can be defined 
as \begin{equation}
    \label{eq:meas_channel}
    \Tilde{\rho}=\mathcal{M}(\rho)=\sum_{a}tr(\rho M_{a})\ketbra{\psi_{a}}{\psi_{a}}.
\end{equation}
In a more general scheme, like the one mentioned in the beginning of the subsection, $\ket{\psi_{a}}$ is a random vector chosen according to a probability distribution. For example, in the current scheme, if the largest eigenvalue of $M_a$ is degenerate, we choose any one of the corresponding eigenvectors with equal probability. 

In the formalism developed in  \cite{Huang_2020}, the channel and its inversion were related to the ensemble of (random) unitary transformations (e.g.~Clifford unitary ensemble). The condition of tomographical completeness depended on the existence of a unitary transformation in the chosen ensemble to distinguish different density matrices \cite{Huang_2020}. However, with our reformulation of the measurement channel, we need to use an informationally complete set POVMs (e.g.~Pauli-6, see Sec. \ref{sec:POVM}). 

In the example of a single qubit measured using the 6 projectors coming from the 3 Pauli matrices i.e. Pauli-6 POVM, the channel and its inverse can be explicitly computed. Similar to the classical shadows built out of random Pauli measurements \cite{Huang_2020}, we get a depolarizing channel i.e. a channel that contracts a pure state (lying on the surface of the Bloch sphere) towards the `center' of the Bloch sphere, namely, the maximally mixed state $\rho=\mathbb{I}_2/{2} $. The inverse (a non-physical map) can be computed, which can map a point inside the Bloch ball to the outside.
\begin{figure}
    \centering
    \includegraphics[width=7cm]{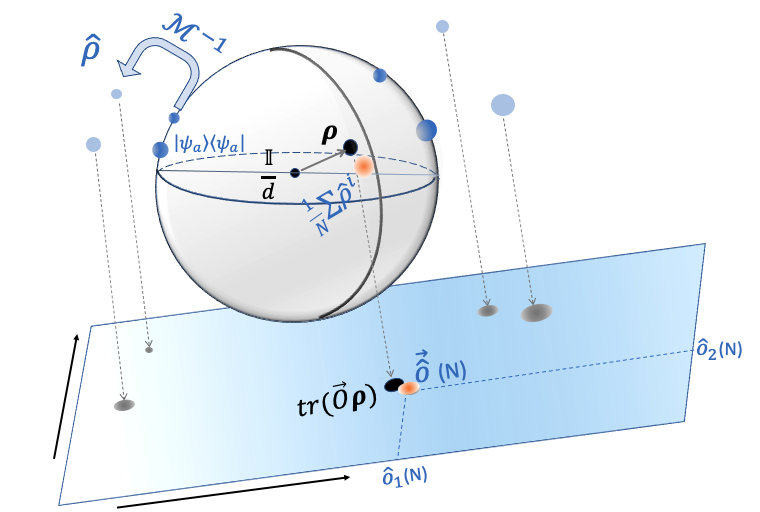}
    \caption{The convex region in the figure is the set of admissible density matrices. We schematically describe the process of forming classical shadows from $N$ copies of $\rho$. For the $i$-th observation with outcome `$a$', the inverse of the channel, $\mathcal{M}^{-1}$, acts on the projectors $\ketbra{\psi_{a}}$ to construct the shadow $\hat{\rho}^{i}$. The sample mean of the shadows cast by $\rho$ i.e. $\frac{1}{N}\sum_{i}\hat{\rho}^{i}$ fluctuates around the true $\rho$ and could be outside the convex region. However, while measuring $L$ $k$-local  observables \cite{Huang_2020} $\vec{O}=(O_{1},\ldots O_{L})$, the convergence of the sample averages $\vec{\hat{o}}(N)$ to the true expected values can be guaranteed with a number of samples $O(\log L)$. See Theorem \ref{Theorem_1}. }
    \label{fig:shadow_vis}
\end{figure}

\textbf{Multi-qubit system}: For local measurements (not necessarily the depolarizing channel), the inverse channel for the $n$-qubit system can be written as \begin{equation}
    \mathcal{M}_{n}^{-1}=\bigotimes_{j=1}^n  \mathcal{M}_{1}^{-1}.
\end{equation}
We can now reformulate the shadows with our overcomplete POVM set and its corresponding channel. For instance, when we work with Pauli-6 POVM, we will get
\begin{equation}
    \label{povm_shadow}
    \hat{\rho} = \bigotimes_{j=1}^n \mathcal{M}_{1}^{-1} (\ketbra{\psi_{a,j}}{\psi_{a,j}})   \qquad  (classical \ shadow),
\end{equation} where 
$ \mathcal{M}_{1}^{-1}(X)=3X-tr(X)\mathbb{I}$  (see Sec.~\ref{sec:Appendix Inv_B}).
Note that the $2^n\times 2^n$ matrix $\hat{\rho}$ need not be constructed explicitly. We just need to store $\ket{\psi_{a,j}}$ for each qubit $j$.

Since the inverted channel $\mathcal{M}^{-1}$ is not physical (it is not completely positive), the $\hat{\rho}$ in Eq.~\eqref{povm_shadow} need not be physical. In other words, there is no guarantee the output of the inverse channel is positive semidefinite. See Fig.~\ref{fig:shadow_vis} for a schematic description. We recover the true density matrix only in expectation. However, if the shadow matrix is forced to be positive semidefinite, we can see how the observations such as fidelity changes (see Sec.~\ref{sec:Appendix A}).

\subsection{Noisy shadow}
Earlier, we defined our measurement channel, Eq.~\eqref{eq:meas_channel}. However, we can also let each of our qubits pass through a previously characterized noise channel $\mathcal{E}_1$ and then take the measurements \cite{koh2020classical}. The combined channel ${M}_{\mathcal{E},1}$ is given by
\begin{equation}
\label{eq:meas_channel_noise}
    \Tilde{\rho}=\mathcal{M}_{\mathcal{E},1}(\rho)=\sum_{a}tr(\mathcal{E}_1(\rho) M_{a})\ketbra{\psi_{a}}{\psi_{a}}.
\end{equation}
We used IC POVMs to ensure that the measurement channel was invertible. As long as the action of the noise channel $\mathcal{E}_1$ itself is invertible, $\mathcal{M}_{\mathcal{E},1}$ is also invertible. We will work with an $n$-qubit noise channel of the form $
    \mathcal{E}_{n}=\bigotimes_{j=1}^n  \mathcal{E}_{1}
$. Thus, we can still write the inverse of the new noisy measurement channel for the $n$-qubit system in terms of the single qubit inverse shadow channel $\mathcal{M}_{\mE,1}^{-1}$: 
\begin{equation}
    \mathcal{M}_{\mE,n}^{-1}=\bigotimes_{j=1}^n  \mathcal{M}_{\mE,1}^{-1}.
\end{equation}

If we choose an amplitude damping channel with damping parameter $\gamma$, one of the Kraus operator representations can be given as \begin{equation}
    \mE_{AD}(\rho)=K_{0}\rho K^{\dagger}_{0} + K_{1}\rho K^{\dagger}_{1}, 
\end{equation}
where $K_{0}=
\begin{bmatrix}
           1 && 0\\
           0 && \sqrt{(1-\gamma)}
         \end{bmatrix}$, $K_{1}=
\begin{bmatrix}
           0 && \sqrt{\gamma}\\
           0 && 0
         \end{bmatrix}$. 

The inverse of the noisy shadow channel  $\mathcal{M}_{{AD}}^{-1}(X)$ is given in Eq.~\eqref{eq:invers_noisy}.~Its action on $\mathbb{I}$ and $\sigma_{x,y,z}$ is given as $\mathcal{M}_{{AD}}^{-1}(\mathbb{I})=\mathbb{I}-\frac{\gamma}{1-\gamma}\sigma_{z}$, $\mathcal{M}_{{AD}}^{-1}(\sigma_{x,y})=\frac{3}{\sqrt{1-\gamma}}\sigma_{x,y}$ and $\mathcal{M}_{{AD}}^{-1}(\sigma_{z})=\frac{3}{1-\gamma}\sigma_{z}$. See Sec.\ref{sec:Appendix Inv_B} for a general description on the inversion of a noisy shadow channel.
Here, we will construct the shadows (noisy) with the following definition:
\begin{equation}
    \label{noisy_shadow}
    \hat{\rho} = \bigotimes_{j=1}^n \mathcal{M}_{AD}^{-1} (\ketbra{\psi_{a,j}}{\psi_{a,j}}). 
\end{equation}
\subsection{Predicting linear functions with classical shadows} 
Using the statistical properties of a single shadow, we can predict linear functions in the unknown state $\rho$ as 
\begin{equation}
     o=tr(O\rho)=\mathbb{E}[\hat{o}], \qquad \textrm{where} \quad \hat{o}=tr(O\hat{\rho}).
\end{equation}
In practice, using an array of shadows (i.e. $N$ snapshots), we can estimate the expectation $o$. Given an array of $N$ independent classical snapshots (each defined as in Eq.~\eqref{povm_shadow}) :
\begin{equation}
\mathsf{S}(\rho;N) = \left\{ \hat{\rho}^{(1)} ,\hat{\rho}^{(1)}, \ldots, \hat{\rho}^{(N)} \right\}.
\end{equation}
The sample mean is $\bar{o} = \frac{1}{N}\sum_{j=1}^N \mathrm{tr} \left( O \hat{\rho}^{(j)} \right).$
This sample mean will fluctuate around the true prediction, with $\mathbb{E}(\bar{o})=o$. 
\subsection{The algorithm and the guarantee of performance}
We want to predict the expected value of multiple $k$-local observables $O_1,\ldots O_L$ based on shadows using the two algorithms below.
\begin{algorithm}[H]
\SetAlgoNoLine
\KwIn{IC POVM with $k$ outcomes, $\rho \in \mathcal{C}^{2^n}$ (N copies of the unknown density matrix)} 
Compute the measurement channel 
$\mathcal{M}_{1}$  and its inverse $\mathcal{M}^{-1}_{1}$  for the chosen IC POVM. (See Sec. \ref{sec:Appendix Inv_B})
\;

\For{$i = 1,\ldots N$} {
Perform measurements using the POVM elements $M_{a}$ to get outcomes $a_{ji} \in \{1,\ldots, k \}$ \;
Construct shadows
$ \hat{\rho}_i = \bigotimes_{j=1}^n \mathcal{M}_{1}^{-1}(\ketbra{\psi_{a_{ji}}}{\psi_{a_{ji}}}) $ (See  Sec. \ref{sec:Measurement_channel}, \ref{Sec:App_Channel} for the general version)

} 

\KwOut{$\hat{\rho}_1,\hat{\rho}_2\ldots \hat{\rho}_N$ }

\caption{Generating Shadows with POVMs}
\label{Alg_1}
\end{algorithm}
 
\begin{algorithm}[H]

\DontPrintSemicolon
\SetAlgoNoLine

\KwIn{A POVM set, N copies of unknown density matrix $\rho$, L different $k$-local Pauli observables $O_{1}, O_{2},\ldots O_{L} $ and error parameters $\epsilon, \delta$} 

Find bounds on the local observables $B(\{O\},\mathcal{M})$. (See Sec.\ref{sec: App Sample Complexity} for details).  \;

Using algorithm.\ref{Alg_1}, collect  $N\geq \frac{B(\{O\},\mathcal{M})\log(\frac{2L}{\delta})}{2\epsilon^{2}}$ shadows.

Compute means $\hat{o}_{i} = \frac{1}{N}\sum_{j=1}^N \mathrm{tr} \left( O_{i} \hat{\rho}^{(j)} \right)$\;

\KwOut{$\bar{o}_{1},\bar{o}_{2}\ldots \bar{o}_{L}$  

}

\caption{Predicting many properties using mean as an estimate}

\end{algorithm}
The existence of the bound is guaranteed by the following theorem.
\begin{theorem}
\label{Theorem_1}
With $N\geq \frac{B(\{O\},\mathcal{M})\log(\frac{2L}{\delta})}{2\epsilon^{2}}$ samples of $\rho$, we can predict $L$ different linear target functions $tr(O_{1}\rho),\  tr(O_{2}\rho),\ldots, tr(O_{L}\rho)$ up to additive error $\epsilon$ with maximum failure probability $\delta$.
\end{theorem}
The constant bound $B(\{O\},\mathcal{M})$ will depend on the measurement channel $\mathcal{M}$ (which depends on the choice of POVM) and on the operator set $\{O\}=\{O_{1},O_{2},..\hspace{1pt},O_{L}\}$). The important thing is that $B(\{O\},\mathcal{M}$ is  bounded for so called $k$-local operators, as defined in \cite{Huang_2020}.  


For instance, if we choose Pauli-6, the bound is given as $\hat{o}^{(j)}_{i}\in[-3^k, 3^k]$, in which case $B(k\text{-local}, \text{Pauli-}6)=4\times9^{k}$ (See Sec.\ref{sec: App Sample Complexity}). 
See section \ref{sec:Measurement_channel} for the algorithm, including the construction of the measurement channel. In the appendix section on sample complexity (Sec. \ref{sec: App Sample Complexity}), the details of the proof is provided.

\section{Numerical Results}
\begin{figure}
    \centering
    {
    \includegraphics[width=.8\linewidth]{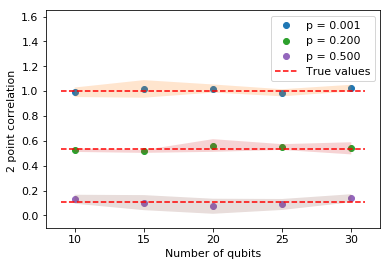}} 
    \caption{Prediction of two-point correlations, $\langle \sigma^{z}_{i}\sigma^{z}_{j} \rangle$, for noisy GHZ target states using classical shadows for Pauli-6, with 1-standard deviation band. The standard deviations are estimated  over ten independent runs, each of which involved $N=5000$ samples.  The parameter $p$, representing the local depolarizing noise strength, is described in Eq.~\eqref{eq:noisy_dep_GHZ}.}
    \label{fig:rough_corr_plot}
\end{figure}

For many quantum systems in Condensed Matter Physics, one of the objects of interest is the two-point correlation function. Two-point correlators could be efficiently estimated using classical shadows based on Pauli-6 POVM. The predictions of two-point functions $\langle \sigma^{Z}_i \sigma^{Z}_j\rangle$ for the GHZ states with varying degree of noise is shown in Fig.~\ref{fig:rough_corr_plot}.

\begin{figure}
\centering

\subfloat[GHZ]{
\includegraphics[width=7cm]{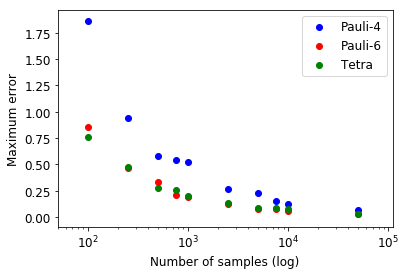}}
\quad \\
\subfloat[Spin down]{
\includegraphics[width=7cm]{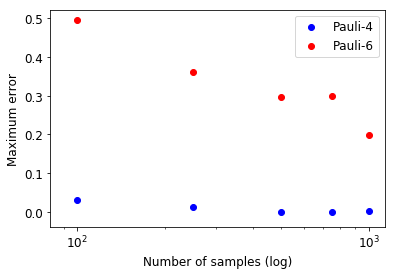}
}
\caption{Maximum error in two-point correlators. (a) Scaling of maximum error among all two-point correlations in 30 qubit pure GHZ state, plotted against different number of samples for different choice of POVMs: Pauli-6, Pauli-4 and tetrahedral.  (b) Scaling of maximum error for all spin down state with Pauli-4 and Pauli-6 POVM. Pauli-4 ensures a much better scaling. See Sec.~\ref{sec: App Sample Complexity} for details. }
    
\label{max_scaling}

\end{figure}

We can write the action of the single qubit depolarizing noise \cite{Nielsen} on an arbitrary $\rho$ written in the Bloch sphere representation:  \begin{equation}
\rho'=(1-\frac{4}{3}p)\rho+\frac{4}{3}p\mathbb{I}.
\label{eq:noisy_dep_GHZ}
\end{equation}
Applying this channel to every qubit, we generate a noise GHZ state \cite{greenberger2007going} from a pure one.
The expected two-point correlations $\langle \sigma^{Z}_i \sigma^{Z}_j\rangle$ varies as $(1-4p/3)^2$ with the noise parameter $p$. \\ 
While predicting multiple $1,\ldots, L$, two-point or $k$-point correlations, we monitor the maximum possible error among all the observables. This measure of error is expected to go down with increasing number of samples. This scaling, as seen in Fig.~\ref{max_scaling}, gives us some idea of the appropriateness of a POVM set for a particular task. 

\subsubsection{1D Transverse Field Ising Model}
\begin{figure}
    \centering
    \subfloat[ $J=h=1$ ]{
    \includegraphics[width=.5\linewidth]{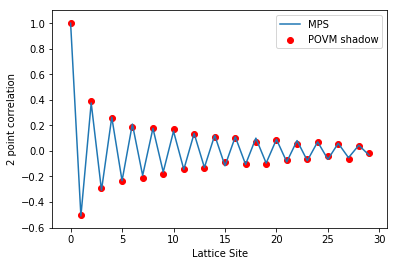} }
    \subfloat[ $J>h$, $J=1$ and $h=0.5$ ]{
    \includegraphics[width=.5\linewidth]{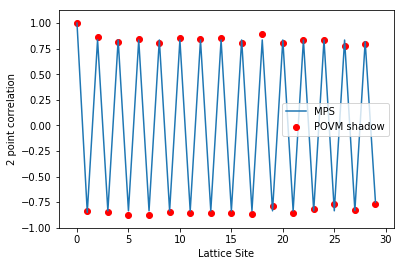} }

    \subfloat[ $J<h$, $J=0.5$ and $h=1$ ]{
    \includegraphics[width=.5\linewidth]{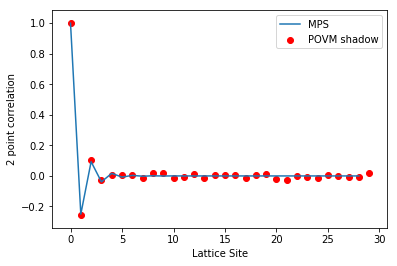} }
    
    \caption{Two-point functions $\langle \sigma^{z}_{0}\sigma^{z}_{i} \rangle$  for ground states of antiferromagnetic 1D tranverse field Ising model using Pauli-6 POVM based shadows and the true value, as computed using matrix product states. The correlations are plotted against the lattice separation. The lattice size is 30 and the number of samples used is 5000. (a) Critical ($J=h$) antiferromagnetic 1D TFIM, showing signatures of power-law correlation. (b) Ordered state ($J>h$), where correlations saturates with increasing lattice separation. (c) The paramagnetic state ($J<h)$, displaying exponential decay of the correlations.} 
  \label{fig:ising_crit_shadow}
\end{figure}

We take antiferromagnetic ($J>0$ in Eq.~ \eqref{eq:1d_tfim} ) transverse field Ising model in 1D: \begin{equation}
\label{eq:1d_tfim}
    H=J\sum_{<ij>}\sigma^{z}_{i}\sigma^{z}_{j} + h\sum_{i}\sigma^{x}.
\end{equation}
The quantum critical point at $h/J=1$ will be exhibited by the power-law decay of the correlations. See Fig.~\ref{fig:ising_crit_shadow} for results in the three regimes: critical, ordered and paramagnetic. The exact numerical correlations are plotted using the matrix product representations of the ground states. \cite{Orus_2014}. In \cite{Carrasquilla_2019} and \cite{Luchnikov_2019}, POVM-based measurements, followed by a neural-network-centric approach for constructing the ground state, and computing the resulting two-point correlations were presented for the same system. 

\subsubsection{1D Disordered Heisenberg Model}
\begin{figure}
\centering
\subfloat[ Exact Diagonalization]{\includegraphics[width=.48\linewidth]{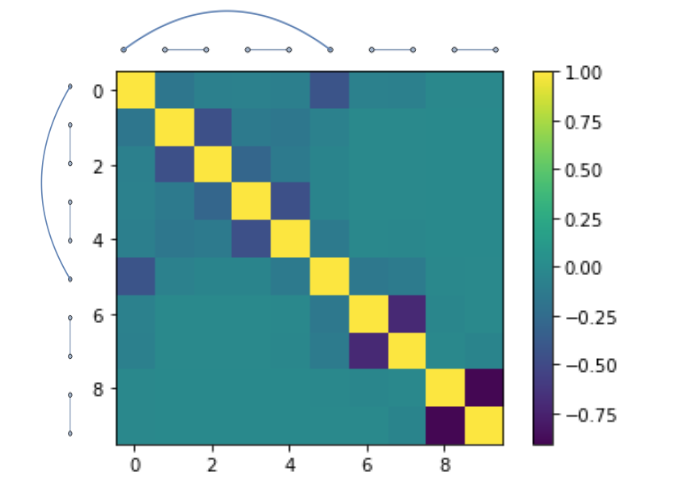}
\label{fig:ed_heis} }
\subfloat[Reconstruction with Shadows]{\includegraphics[width=.48\linewidth]{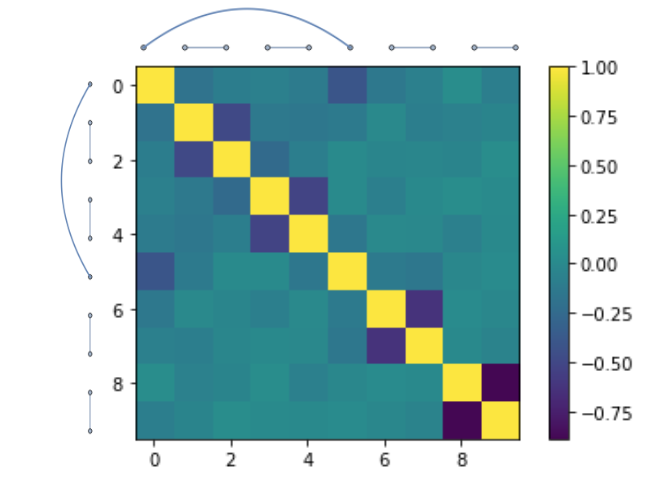}
\label{fig:shad_heis} }
\caption{Two-point functions $\langle \sigma^{z}_{i}\sigma^{z}_{j} \rangle$ for ground states of disordered 1D Heisenberg spin chain with length=10 and open boundary conditions. (a) Exact diagonalization results. (b) Results from using Pauli-6 POVM based shadows using 5000 samples. } 
\label{fig:heiss_crit_shadow}
\end{figure}

The Hamiltonian for the 1D disordered Heisenberg model is given by,  \begin{equation}
\label{eq:heis}
\hat H = -\frac{1}{2} \sum_{j=1}^{N} (J^{x}_{j} \sigma_j^x \sigma_{j+1}^x + J^{y}_{j} \sigma_j^y \sigma_{j+1}^y + J^{z}_{j} \sigma_j^z \sigma_{j+1}^z + h\sigma_j^{z}).
\end{equation}
The properties of spin-$\frac{1}{2}$ antiferromagnetic chains with various types of random exchange coupling has been studied in an exact decimation renormalization-group (strong-disorder) schemes, some of which involve generalization or modifications of the scheme introduced by Dasgupta and Ma \cite{1980PhRvB..22.1305D}. The numerical studies done by R.N. Bhatt and P.A. Lee \cite{Bhatt_Lee} indicate that the system could be in a random-singlet phase. In such a phase, each spin is paired with another spin that may be far away on the lattice. We perform exact diagonalization, obtain the ground state and then compute two-point quantum correlations. The 2d plot of the correlation matrix will also inform us about the locations of the singlet formations in the chain. We can also reconstruct these behavior of a ground-state corresponding to one particular disorder realization of the XXZ-Heisenberg model Eq.~\eqref{eq:heis} ($J_{x}=J_{y}=2J_{z}$, $h=0$) with sufficient number of shadows. See Fig.~\ref{fig:heiss_crit_shadow}, where the singlet formations are indicated by the schematics drawn on the axes of the matrix visualization plots and the results from the two methods are compared. \\\\   

\subsection{Exploring quantum fidelity}
\label{sec:Appendix A}

In our approach to construct shadows using local POVMs, we ensure prediction of local observables. However, we can also explore non-local observables such as fidelity.
Using sample mean as an estimator, we can construct a hypothesis state ($\sigma$): \begin{equation}
    \sigma=\frac{1}{N}\sum^{N}_{i}\hat{\rho}.
\end{equation}

When our target state is pure, we can rewrite quantum fidelity as a linear prediction with our target observable given as $O=\ketbra{\psi}{\psi}$. Starting from this definition of quantum fidelity i.e. $F_{Q}(\rho,\sigma)=(tr(\sqrt{\sqrt{\rho}\sigma\sqrt{\rho} }))^{2}$, using $\rho=\ketbra{\psi}{\psi}$ we get $(tr\sqrt{( \bra{\psi}\sigma\ket{\psi} \ketbra{\psi}{\psi} )} )^{2}$.  Further simplification of the fidelity gives us $$ F_{Q}(\rho,\sigma) = \bra{\psi}\sigma\ket{\psi} (tr(\sqrt{\ketbra{\psi}{\psi}}))^{2}=\bra{\psi}\sigma\ket{\psi}=tr(\sigma O). $$ 

\begin{figure}
    \centering
    \includegraphics[width=6.5cm]{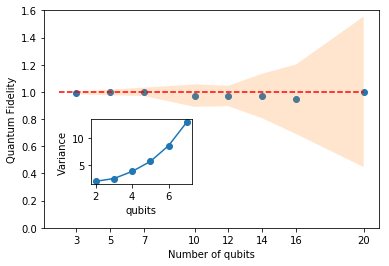}
    \caption{Quantum fidelity predicted for the pure GHZ state using sample mean of shadows constructed on $10^4$ samples. The shaded regions are the standard deviation over ten independent runs. The inset shows the scaling of the variance of  fidelity which grows exponentially with number of qubits.}
    \label{fig:fidelity_no_proj}
\end{figure}

The measure $tr(\sigma O)$ is equivalent to fidelity, only when the latter is defined, i.e. when $\sigma \succeq 0$.  That property is likely to hold only when the number of samples is large. We expect $tr(\sigma O)$ to fluctuate around its mean value 1, as seen in Fig.~\ref{fig:fidelity_no_proj}, even when the typical $\sigma$ is not a physical state, meaning it is not positive semidefinite. Also, the fluctuation around this mean keeps on growing exponentially with the number of qubits (see Fig.~\ref{fig:fidelity_no_proj}). This growth cannot be dealt with even by the median of means (MoM) procedure \cite{Huang_2020} within the shadow formalism. Numerical computations using MoM also show no advantage over sample means here.

Hence, we need a procedure to find the `closest' physical state to $\sigma$. The trace condition $tr(\hat{\sigma})=1$ ensures that once $\hat{\sigma}\nsucc 0$, some of the eigenvalues will be greater than 1 to compensate for the negative eigenvalues. Thus, we cannot just throw away the negative eigenvalues, as would be done for projecting a Hermitian matrix to the space of positive semidefinite matrices.

We define the the convex set of physical states to be $C=\{\rho|\rho \succeq 0, tr(\rho)=1\}$. Our nonlinear projection to $C$ is
\begin{equation}
\Pi_C(\sigma)=\argmin_{\rho\in C} tr((\rho-\sigma)^2).
\end{equation}

\begin{figure}
\centering
\includegraphics[width=6cm]{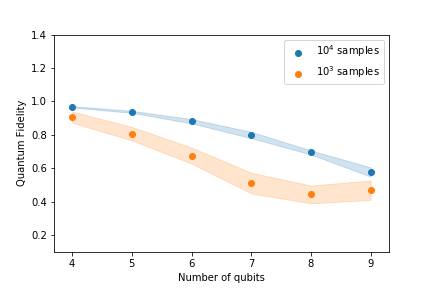}
\caption {Quantum fidelity of the projected shadows (onto the physical positive definite space) with the noiseless GHZ state. As we increase the number of samples, from $10^3$ to $10^4$, the quantum fidelity improves. The shaded regions indicate 1-standard deviation bands, estimated over ten independent runs.}
\label{fig:projected_shadows}
\end{figure}

We achieve this by diagonalizing $\sigma$, projecting the eigenvalues~$\lambda_i$ of~$\sigma$ onto a canonical simplex $\Delta = \{(\lambda^{p}_1, \dots, \lambda^{p}_D) \mid \lambda^{p}_i \ge 0, \sum_{i=1}^D \lambda^{p}_i = 1\}$, using the recipe from Ref.~\cite{proj_wang}, while leaving the eigenvectors untouched. Here, $D=2^{n}$ where $n$ is the total number of qubits. The projected state is a  biased estimator.  We can hope that the price paid by accepting some bias comes with the benefit of reduced variance. This expectation seems to be born out in Fig.~\ref{fig:projected_shadows}. However, as number of qubits increase, the bias itself reduces fidelity. To compensate this effect, we need larger sample sizes ($N$). Fig.~\ref{fig:projected_shadows} shows all these trends.

\section{Discussions}
We provide an approach to predict expectations of local observables without having to apply random unitary transformations, which sometimes require complex circuits of its own, and can become a practical bottleneck.
 We show that this can rather be done using an IC  POVM.  For illustrations, we show faithful reconstruction properties of low energy states coming from different many body Hamiltonians relevant to near-term applications of quantum devices. When we have additional information about the possible noisy channels we also adapt the shadow channel as a composition of the noise channel and the measurement channel. The invertibility becomes straightforward in the proposed framework. We also comment on why the mean as an estimator is sufficient throughout our discussion. And as long as we are dealing with local observables, we can provide efficient sample complexity using Hoeffding's inequality directly. 
 
 We provided instances where the choice of POVM impacts the sample complexity for predicting 2-point correlators in certain quantum states for fixed maximum error. We noted that the different POVMs work better for different states. It is an exciting endeavour to understand which sets of POVM would be ideal for different classes of quantum states and observables.
 
 Although, an exploration, we attempt to reconstruct fidelity using the locally built shadows and show that we cannot benefit from median of means as an estimator, since variance of fidelity becomes exponential in number of qubits. Additionally, when presented with few samples we raise the issue of unphysical i.e. not positive semidefinite $\hat{\rho}$ and then provide a projection tecnique, similar to \cite{Struchalin_2021}, to estimate fidelity. Unfortunately, the estimator no longer remains unbiased. Addressing this issue would require methods to deal with non-local observables.
 
We did not provide an effective analog of the global Clifford unitary transformation-based method in \cite{Huang_2020}. There has been work which provides description of global alternatives using stabilizer states \cite{Struchalin_2021}. Whether there can be a scheme based on such states that is competitive with the classical shadows method \cite{Huang_2020} remains to be seen.
 
The use of generalized measurement to unambiguously discriminate non-orthogonal states with lower failure probability is well known  \cite{Barnett:09,Nielsen, Chefles_2000}. Efficient prediction of expectations of local observables combined with the generalized measurement scheme to obtain the shadows can be used as an optimal framework in the discrimination of non-orthogonal states. In the future, it is a promising direction of exploration.  
\section*{Acknowledgement}
We would like to thank Shagesh Sridharan, James Stokes, Miles Stoudenmire for insightful discussions. 
\appendix

\section{Appendix}

\label{Sec:App_Channel}
\subsection{The Measurement Channel for Pauli-6}


We can take the simple rank-1 Pauli-6 POVMs to see the action of a measurement channel: \begin{equation}
    \label{eq:q1u_pauli}
    \Tilde{\rho}=\mathcal{M}(\rho)=\sum_{a}tr\big(\frac{1}{2}(\mathbb{I}+\mathbf{r}.\pmb{\sigma}) M_{a}\big)\ketbra{\psi_{a}}{\psi_{a}},
\end{equation}
where we use the Bloch representation $\rho=\tfrac{1}{2}(\mathbb{I}+\mathbf{r}.\pmb{\sigma})$.

The contribution of the first two POVM elements of Pauli-6 only gets contribution from $\mathbb{I}$ and $r_z\sigma_z$, generating 
\[
    tr(\frac{1}{2}(\mathbb{I}+r_{z}\sigma_{z}) M_0)\ketbra{0}{0} +tr(\frac{1}{2}(\mathbb{I}+r_{z}\sigma_{z}) M_1)\ketbra{1}{1}.
\]

Using $M_0=\frac{1}{3}\times\ketbra{0}{0}$ and  $M_1=\frac{1}{3}\times\ketbra{1}{1}$ this expression becomes: $\frac{1}{6}\mathbb{I}+\frac{1}{3}r_{z}\sigma_{z}$
Following similar steps for pairs $M_{2},M_{3}$ and $M_{4},M_{5}$, we get that: $$ \mathcal{M}(\rho)=\mathcal{M}\big(\frac{1}{2}(\mathbb{I}+\mathbf{r}.\pmb{\sigma})\big)=\frac{1}{2}(\mathbb{I}+\frac{1}{3}\mathbf{r}.\pmb{\sigma}),  $$ making $\mathcal{M}$ a depolarizing channel.
\subsection{Inverse of the measurement channel}
\label{sec:Appendix Inv_B}
Given any single qubit channel, the inverse can be easily computed using the\textit{ Bloch-sphere} representation. We can write any 2 dimensional  (single qubit) quantum operation ($X$) as  
 $X=(x_{0}\mathbb{I}+\vec{r}.\vec{\sigma})$. Any arbitrary trace-preserving quantum operation is given as $\mathcal{E}(X)=(x_{0}\mathbb{I}+\vec{r}^{\hspace{2pt}\prime}.\vec{\sigma})$. The map $\vec{r}  \xrightarrow{\mathcal{E}} \vec{r}^{\hspace{2pt}\prime}$ is equivalent to, \begin{equation}
 \vec{r}^{\hspace{2pt}\prime}=  T\vec{r}+x_{0}\vec{c}, \quad T_{i,j} =\frac{1}{2}tr(\sigma_{i}\mathcal{E}(\sigma_{j}) ).  \end{equation} The components of displacement ($\vec{c}$) is given as $ c_{i} =\frac{1}{2}tr(\sigma_{i}\mathcal{E}(\mathbb{I}) )$. The \textit{affine map} between the Bloch sphere and itself is given by $T$, and its meaning is understood better by doing a singular value decomposition i.e. $T=O_{1}DO'_{2}$ where $O_{1},O_{2}$ are orthogonal matrices. The singular values capture the deformation of the Bloch sphere about its principal axes.
A superoperator $\hat{T}_{4\times4}$ can be defined as  \begin{align}
    \hat{T}\begin{bmatrix}
           x_{0} \\
           \vec{r}
         \end{bmatrix}=\begin{bmatrix}
           x_{0} \\
           \vec{r}^{\hspace{2pt}\prime}
         \end{bmatrix},  \quad \hat{T}_{4\times4} &= \begin{bmatrix}
           1 && 0\\
           \vec{c} && \small{T_{3\times3}}
         \end{bmatrix}.
  \end{align}
Computing inverse of the channel is equivalent to writing $(x_{0},\vec{r})$ from $(x_{0},\vec{r}^{\hspace{2pt}\prime})$ i.e. computing $\hat{T}^{-1}$.

In the main text, the Pauli measurement channel ($\mathcal{E}=\mathcal{M}_{1}$) turns out to be a depolarizing channel, and its inverse that acts on the local qubit is given as \begin{equation*}
    \mathcal{M}_{1}^{-1}(X)=3X-tr(X) \mathbb{I}.
\end{equation*}

We take a more general example following our definition of a measurement channel:  \[
    \label{eq:meas_channel1}
    \Tilde{\rho}=\mathcal{M}(\rho)=\sum_{a}tr(\rho M_{a})\ketbra{\psi_{a}}{\psi_{a}}.
\]
If a particular POVM element is not  rank one,  $\ket{\psi_{a}}$ can be taken as the eigenvector corresponding to the highest eigenvalue of $M_{a}$. For Pauli-4, except for $M_3=\frac{1}{3}(\ketbra{1}{1}+ \ketbra{-}{-}+ \ketbra{r}{r})=\mathbb{I}$, all other elements are rank-1. Since $M_{3}$ is rank-2, when the outcome is $3$, we take the eigenvector $\ket{t}$ corresponding to eigenvalue $\frac{1}{2}(1+\frac{1}{\sqrt{3}})$ instead of the other corresponding to $\frac{1}{2}(1-\frac{1}{\sqrt{3}})$. Rewriting Eq.~\eqref{eq:meas_channel1}, we get 
\begin{equation}
\begin{split}
    \Tilde{\rho}=\mathcal{M}_{1}(\rho)=tr(\rho M_{0})\ketbra{0}{0}+tr(\rho M_{1})\ketbra{+}{+} \\ + tr(\rho M_{2})\ketbra{l}{l}+M_{3})\ketbra{t}{t}.
\end{split}
\end{equation}
The inverse of the channel can be written as  
\begin{equation}
\begin{split}
    \mathcal{M}_{1}^{-1}(X)=6X-(\frac{x_{s}}{\sqrt{3}} - x_{0}(\sqrt{3}-1))(\sum_{i}\sigma_{i})-5x_{0}\mathbb{I}\\ \quad x_{0}=\frac{tr(X)}{2}, \hspace{3pt}  x_{s}=\frac{3(tr(X(\sum_{i}\sigma_{i}))+ (\sqrt{3}-1)x_{0} )}{\sqrt{3}+1}.
\end{split} 
\end{equation}
When we are working with a known noise channel  $\mathcal{E}$,  the inverse is given as $\hat{T}_{\mathcal{M}_{\mathcal{E}}}^{-1}=\hat{T}_{\mathcal{E}}^{-1}\hat{T}_{\mathcal{M}}^{-1}$.
If we choose an amplitude damping channel with a damping parameter $\gamma$, the inverse can be given as
\begin{equation}
\label{eq:invers_noisy}
\begin{split}
\mathcal{M}_{\mathcal{E}}^{-1}(X)=\frac{3(1-\frac{1}{2}tr(X\sigma_{z}))}{\sqrt{(1-\gamma)}}X+\frac{3tr(\sigma_{z}X)}{2(1-\gamma)}X\\+(\frac{1}{2}-\frac{3}{2\sqrt{(1-\gamma)}})tr(X)\mathbb{I}+\frac{\gamma tr(X)}{2(\gamma-1)}\sigma_{z}.
\end{split}
\end{equation}
\subsection{Sample complexity}
\label{sec: App Sample Complexity}

\subsubsection{Variance of the Estimate for a Single Observable}
Given an array of $N$ independent, classical snapshots (each defined as Eq.~\eqref{povm_shadow}) :
\begin{equation}
\mathsf{S}(\rho;N) = \left\{ \hat{\rho}^{(1)} ,\hat{\rho}^{(1)}, \ldots, \hat{\rho}^{(N)} \right\}.
\end{equation}
The sample mean is $\hat{o} = \frac{1}{N}\sum_{j=1}^N \mathrm{tr} \left( O \hat{\rho}^{(j)} \right).$
The bound on probability of deviation of the sample mean is given by Chebyshev's inequality:
\begin{equation}
\label{cheby}
  Pr(\abs{\hat{o}-\mathbb{E} \left[ \hat{o} \right]} \geq \epsilon ) \leq \frac{Var[\hat{o}]}{\epsilon ^{2}} 
\end{equation}
$\mathbb{E} \left[ \hat{o} \right] = \mathrm{tr} \left( O \rho \right)$ where $\rho$ is the true density matrix.
Fluctuations of $\hat{o}$ around this desired expectation are controlled by the variance. $Var[\hat{o}]=\frac{1}{N} Var[tr(O\hat{\rho}^{(j)})]=\frac{Var[o^{(j)}]}{N}$. However, since the classical shadows are unit trace by construction, the variance depends only on the trace-less part of the observable i.e. $O_{0}=O-\frac{tr(O)}{2^{n}}\mathbb{I}$. The minimum number of samples needed to assure a maximum failure probability ($\delta$) using   Eq.~\eqref{cheby} is   \begin{equation}
    N \geq \frac{Var[o^{(j)}]}{\epsilon^{2} \delta}.
\end{equation}

\subsubsection{Dependence on POVM}
Given a measurement channel and an observable, we can bound the variance of its estimator, using familiar maneuvers with superoperators  \cite{Huang_2020},
\begin{flalign*}
Var[o^{(j)}]=\mathbb{E}((o^{(j)})^2)-(\mathbb{E}(o^{(j)}))^{2}\leq\mathbb{E}((o^{(j)})^2)\\=
    \sum_{a_{1},., a_{n}}Pr(a_{1}..a_{n})\bra{a_{1},., a_{n}}[\mathcal{M}^{-1}_{n}]^{\dagger}(O_{0})\ket{a_{1},., a_{n}}^{2} \\\text{where} \quad Pr(a_{1},..,a_{n})=tr(\rho M^{a_{1}}\otimes M^{a_{1}}..\otimes M^{a_{n}}).
\end{flalign*}
We broadly define a $k$-local Pauli-observable as an operator which acts nontrivially only on $k$ qubits. Traceless $k$ local operators can be expressed as linear conbination of tensor products of indentity matrices and $k$ or less Pauli matrices. Hence, we need to focus only on special class of $k$-local operators. Denoting $P_{i}$ as one of the Pauli matrices acting on the $i$th qubit,  we focus on of tenor products like $O_{0}=P_{1}\otimes P_{2}\otimes..\otimes P_{k}\otimes \mathbb{I}^{\otimes (n-k)}$, where, without loss of generality, we assume that the operator acts non-trivially on only the first $k$ qubits.

For Pauli-6 POVM, the inverse of the measurement channel is a self-adjoint map, and thus one can verify its action as: \begin{align*}
[\mathcal{M}_{1}^{-1}]^\dagger (P_{\alpha}) = \mathcal{M}_{1}^{-1} (P_{\alpha}) = 3P_{\alpha}, 
\end{align*}
where $P_{\alpha}$ denotes a Pauli matrix and $[\mathcal{M}_{1}^{-1}]^\dagger (\mathbb{I}) = \mathcal{M}_{1}^{-1} (\mathbb{I}) = \mathbb{I}$. 
 Given a  $k$-local observable, we can further compute the bound on variance: \begin{align*}
    Var[o^{(j)}] \leq \sum_{a_{1},., a_{n}}Pr(a_{1}..a_{n})\prod^{k}_{i=1}\bra{a_{i}}3P_{i}\ket{a_{i}}^{2} \\
    =\sum_{a_{1},., a_{n}}tr(\rho M^{a_{1}}\otimes M^{a_{2}}..\otimes M^{a_{n}})\prod^{k}_{i=1}\bra{a_{i}}3P_{i}\ket{a_{i}}^{2} \\
    =tr[\rho \sum_{a_{1},., a_{n}} (  M^{a_{1}}\otimes M^{a_{1}}..\otimes M^{a_{n}} \prod^{k}_{i=1}\bra{a_{i}}3P_{i}\ket{a_{i}})^{2} ] \\
    =tr[\rho \bigotimes^{k}_{i=1} \underbrace{\sum_{a_{i}}M^{a_{i}}\bra{a_{i}}3P_{i}\ket{a_{i}}^{2}}_{3\mathbb{I}}\bigotimes \mathbb{I}^{\otimes(n-k)} ]
    =3^{k}.
\end{align*}
Now, we take up Pauli-4 POVM. One can verify the action of $[\mathcal{M}_{1}^{-1}]^\dagger$ as:
\begin{align*}
[\mathcal{M}_{1}^{-1}]^\dagger (P_{\alpha}) = (2-\sqrt{3})\mathbb{I} + (3+\sqrt{3})P_{\alpha} - \sum_{\beta \neq \alpha}(3-\sqrt{3})P_{\beta}, 
\end{align*}
where $P_{\alpha}$ denotes a Pauli matrix. Using the fact that $\mathcal{M}_{1}^{-1}$ is a trace preserving map, one can say its adjoint has to be unital $i.e.$ $[\mathcal{M}_{1}^{-1}]^\dagger (\mathbb{I}) = \mathbb{I}$. Given a $k$-local Pauli-observable, one can again compute the bound on variance:
\begin{eqnarray*}
    Var[o^{(j)}] \leq \sum_{a_{1},., a_{n}}Pr(a_{1}..a_{n})\prod^{k}_{i=1}\bra{a_{i}}[\mathcal{M}_{1}^{-1}]^\dagger(P_{i})\ket{a_{i}}^{2}\\
    =\sum_{a_{1},., a_{n}}tr(\rho M^{a_{1}}\otimes M^{a_{2}}..\otimes M^{a_{n}})\prod^{k}_{i=1}\bra{a_{i}}[\mathcal{M}_{1}^{-1}]^\dagger(P_{i})\ket{a_{i}}^{2}\\
    =tr[\rho \sum_{a_{1},., a_{n}} (  M^{a_{1}}\otimes M^{a_{1}}..\otimes M^{a_{n}} \prod^{k}_{i=1}\bra{a_{i}}[\mathcal{M}_{1}^{-1}]^\dagger(P_{i})\ket{a_{i}}^{2}] \\
    =tr[\rho \bigotimes^{k}_{i=1} \underbrace{\sum_{a_{i}}M^{a_{i}}\bra{a_{i}}[\mathcal{M}_{1}^{-1}]^\dagger(P_{i})\ket{a_{i}}^{2}}_{5\mathbb{I}+4P_{i}}\bigotimes \mathbb{I}^{\otimes(n-k)} ] \\
    =tr[\rho \bigotimes^{k}_{i=1}{(5\mathbb{I}+4P_{i})}\bigotimes \mathbb{I}^{\otimes(n-k)} ].
\end{eqnarray*}

Clearly, the above bound on variance is dependent on the state $\rho$, unlike the bound we obtained using Pauli-6 POVM. Since $\rho$ is a density matrix and the operator $5\mathbb{I}+4P_{i}$ is a PSD operator, one gets the minimum value for the bound when $\rho$ is of the form: 
\begin{equation}
    \rho = \left(\bigotimes^{k}_{i=1}{\ketbra {p_{i}}}\right) \bigotimes \tilde{\rho}_{n-k},
\end{equation}
where $\ketbra {p_{i}}$ is the projector into the eigenvector corresponding to the lowest eigenvalue of the operator $5\mathbb{I}+4P_{i}$, and $\tilde{\rho}_{n-k}$ is a valid density matrix in the Hilbert space of $n-k$ qubits on which the $k$-local Pauli-observable acts trivially. For the above $\rho$, it is simple to verify that the value of the variance bound is 1 (independent of $k$). Thus, for example, if the unknown state $\rho$ is the all spin down state, then Pauli-4 POVM works better than Pauli-6 POVM in predicting two-point correlators  $\langle \sigma^{Z}_i \sigma^{Z}_j\rangle$, since the variance is higher in the latter.

\subsubsection{Improved Bound Using Hoeffding's Inequality}
Furthermore, we can use Hoeffding's inequality to provide theoretical bounds when we are dealing with $k$-local Pauli observable, since we are working with bounded random variables. If  $\hat{o} = \frac{1}{N}\sum_{j=1}^N \mathrm{tr} \left( O \hat{\rho}^{(j)} \right)$, $\hat{o}^{(j)}\in [a,b]$ for all $j$, where $-\infty<a \leq b \leq \infty$, we can write,
\begin{equation}
\label{hoeff}
  Pr(\abs{\hat{o}-\mathbb{E} \left[ \hat{o} \right]} \geq \epsilon ) \leq 2 e^{(\frac{-2N\large{\epsilon}^{2}}{(b-a)^2})}. 
\end{equation}
The minimum number of samples needed to assure a maximum failure probability ($\delta$) among all the observables using   Eq.~\eqref{hoeff} is  \begin{equation}
    N\geq\log(\frac{2}{\delta}) \frac{(b-a)^{2}}{2\epsilon^{2}}. 
\end{equation}
$B(\{O\},\mathcal{M})=(b-a)^{2}$ depends on the locality of the observable and the maximum eigenvalue $\lambda_{max}$ of the inverse channel acting on the observable. The bound on random variable $\hat{o}^{(j)}$ can be found as the range of the Rayleigh quotient of the inverse of the measurement channel, acting on the observable over all possible states. For instance, if we choose Pauli-6, the bounds can be shown to lie within $\hat{o}^{(j)}\in[-3^k, 3^k]$ in which case $B(k\text{-local}, \text{Pauli-}6)=4\times9^{k}$.
Using the action of $[\mathcal{M}^{-1}]^\dagger$, one can verify that the value of the random variable $tr([\mathcal{M}^{-1}]^\dagger(P_{i}) \ketbra {a_{i}})$ belongs to the set $\{ 5,-1\}$ for any Pauli matrix $P_{i}$ when $\ketbra {a_{i}})$ is the inferred state for Pauli-4. Thus, the random variable $\hat{o}$ is contained in the range $\{ -5^{k}, 5^{k} \}$ which is exponential on the locality rather than the number of qubits.

\subsubsection{The Guarantee of Performance for Multiple Observables}
If we have $L$ different $k$-local Pauli observables $\hat{o}_{1},\ldots, \hat{o}_{i},\ldots ,\hat{o}_{L} $ with the sample mean corresponding to the observable $i$ defined as $\hat{o}_{i} = \frac{1}{N}\sum_{j=1}^N \mathrm{tr} \left( O_{i} \hat{\rho}^{(j)} \right)$.
If  $\hat{o}_{i} = \frac{1}{N}\sum_{j=1}^N \mathrm{tr} \left( O_{i} \hat{\rho}^{(j)} \right)$, $\hat{o}^{(j)}_{i}\in [a,b]$ for all $j$, where $-\infty<a \leq b < \infty$, we can combine the union bound with Hoeffding's inequality to write
\begin{equation}
\label{hoeff}
  Pr(\max\limits_{\substack{1\leq i \leq L}}\abs{\hat{o}_{i}-\mathbb{E} \left[ \hat{o}_{i} \right]} \geq \epsilon ) \leq 2L e^{(\frac{-2N\large{\epsilon}^{2}}{(b-a)^2})}. 
\end{equation}
The minimum number of samples needed to assure a maximum failure probability ($\delta$) among all the observables using   Eq.~\eqref{hoeff} is  \begin{equation}
    N\geq\log(\frac{2L}{\delta}) \frac{(b-a)^{2}}{2\epsilon^{2}}. 
\end{equation}
The scaling is logarithmic in the number of observables $L$, instead of linear behavior we get using Chebyshev's inequality. We do not need to use MoM  procedure \cite{Huang_2020}, which would have been necessary if we were dealing with estimate distributions with long tails (unlike the bounded estimates for $k$-local Pauli observables).
\subsection{Numerical computations}
The computations for GHZ states has been carried out using MPS matrix product state (MPS) representations for noiseless states and matrix product operator (MPO) representations for the noisy states. The details on the simulation of mixed states using MPOs has been shown in \cite{PhysRevLett.93.207204,Carrasquilla_2019}.
The data-sets corresponding to the ground states of spin Hamiltonians such as the Transverse-field Ising model have been generated using the density matrix renormalization group (DMRG). The library used is mpnum (a matrix product represenation library for Python). \cite{Suess2017}.
Given a particular spin model, with the Hamiltonian expressed as a MPO, the
DMRG algorithm attempts to find the optimal MPS with the lowest energy. However, for the visualization of correlations in the disordered 1d Heisenberg spin chain, the computations are made without using the DMRG framework. This is because the number of sites were low for this particular numerical experiment. 

\bibliographystyle{plain}
\bibliography{Papers}

\begin{thebibliography}{10}

\bibitem{Aaronson_2007}
Scott Aaronson.
\newblock The learnability of quantum states.
\newblock {\em Proceedings of the Royal Society A: Mathematical, Physical and
  Engineering Sciences}, 463(2088):3089–3114, Sep 2007.

\bibitem{aaronson2018shadow}
Scott Aaronson.
\newblock Shadow tomography of quantum states, 2018.

\bibitem{Barnett:09}
Stephen~M. Barnett and Sarah Croke.
\newblock Quantum state discrimination.
\newblock {\em Adv. Opt. Photon.}, 1(2):238--278, Apr 2009.

\bibitem{Bhatt_Lee}
R.~N. Bhatt and P.~A. Lee.
\newblock Scaling studies of highly disordered spin-\textonehalf{}
  antiferromagnetic systems.
\newblock {\em Phys. Rev. Lett.}, 48:344--347, Feb 1982.

\bibitem{Carrasquilla_2019}
Juan Carrasquilla, Giacomo Torlai, Roger~G. Melko, and Leandro Aolita.
\newblock Reconstructing quantum states with generative models.
\newblock {\em Nature Machine Intelligence}, 1(3):155–161, Mar 2019.

\bibitem{Chefles_2000}
Anthony Chefles.
\newblock Quantum state discrimination.
\newblock {\em Contemporary Physics}, 41(6):401–424, Nov 2000.

\bibitem{1980PhRvB..22.1305D}
Chandan {Dasgupta} and Shang-Keng {Ma}.
\newblock {Low-temperature properties of the random Heisenberg
  antiferromagnetic chain}.
\newblock {\em \prb}, 22(3):1305--1319, August 1980.

\bibitem{dimic2018single}
Aleksandra Dimi{\'c} and Borivoje Daki{\'c}.
\newblock Single-copy entanglement detection.
\newblock {\em npj Quantum Information}, 4(1):1--8, 2018.

\bibitem{greenberger2007going}
Daniel~M. Greenberger, Michael~A. Horne, and Anton Zeilinger.
\newblock Going beyond {Bell's} theorem.
\newblock eprint arXiv: 0712.0921, 2007.

\bibitem{Huang_2020}
Hsin-Yuan Huang, Richard Kueng, and John Preskill.
\newblock Predicting many properties of a quantum system from very few
  measurements.
\newblock {\em Nature Physics}, Jun 2020.

\bibitem{koh2020classical}
Dax~Enshan Koh and Sabee Grewal.
\newblock Classical shadows with noise.
\newblock e-print arXiv:2011.11580, 2020.

\bibitem{Luchnikov_2019}
Ilia~A. Luchnikov, Alexander Ryzhov, Pieter-Jan Stas, Sergey~N. Filippov, and
  Henni Ouerdane.
\newblock Variational autoencoder reconstruction of complex many-body physics.
\newblock {\em Entropy}, 21(11):1091, Nov 2019.

\bibitem{morris2019selective}
Joshua Morris and Borivoje Daki{\'c}.
\newblock Selective quantum state tomography.
\newblock {\em arXiv preprint arXiv:1909.05880}, 2019.

\bibitem{Nielsen}
M.~A Nielsen and I.~L. Chuang.
\newblock {\em Quantum Computation and Quantum Information: 10th Anniversary
  Edition}.
\newblock Cambridge University Press, 10 edition, 2011.

\bibitem{Orus_2014}
Román Orús.
\newblock A practical introduction to tensor networks: Matrix product states
  and projected entangled pair states.
\newblock {\em Annals of Physics}, 349:117–158, Oct 2014.

\bibitem{Renes_2004}
Joseph~M. Renes, Robin Blume-Kohout, A.~J. Scott, and Carlton~M. Caves.
\newblock Symmetric informationally complete quantum measurements.
\newblock {\em Journal of Mathematical Physics}, 45(6):2171–2180, Jun 2004.

\bibitem{Struchalin_2021}
G.I. Struchalin, Ya.~A. Zagorovskii, E.V. Kovlakov, S.S. Straupe, and S.P.
  Kulik.
\newblock Experimental estimation of quantum state properties from classical
  shadows.
\newblock {\em PRX Quantum}, 2(1), Jan 2021.

\bibitem{Suess2017}
Daniel Suess and Milan Holz\"apfel.
\newblock {mpnum}: A matrix product representation library for {Python}.
\newblock {\em Journal of Open Source Software}, 2(20):465, 2017.

\bibitem{PhysRevA.86.062107}
Gelo Noel~M. Tabia.
\newblock Experimental scheme for qubit and qutrit symmetric informationally
  complete positive operator-valued measurements using multiport devices.
\newblock {\em Phys. Rev. A}, 86:062107, Dec 2012.

\bibitem{Torlai_2018}
Giacomo Torlai, Guglielmo Mazzola, Juan Carrasquilla, Matthias Troyer, Roger
  Melko, and Giuseppe Carleo.
\newblock Neural-network quantum state tomography.
\newblock {\em Nature Physics}, 14(5):447–450, Feb 2018.

\bibitem{PAC}
L.~G. Valiant.
\newblock A theory of the learnable.
\newblock {\em Communications of the ACM}, 1984.

\bibitem{PhysRevLett.93.207204}
F.~Verstraete, J.~J. Garc\'{\i}a-Ripoll, and J.~I. Cirac.
\newblock Matrix product density operators: Simulation of finite-temperature
  and dissipative systems.
\newblock {\em Phys. Rev. Lett.}, 93:207204, Nov 2004.

\bibitem{proj_wang}
Weiran Wang and Miguel~{\'{A}}. Carreira{-}Perpi{\~{n}}{\'{a}}n.
\newblock Projection onto the probability simplex: An efficient algorithm with
  a simple proof, and an application.
\newblock {\em CoRR}, abs/1309.1541, 2013.

\end{thebibliography}

\end{document}